\documentclass[11pt]{article}
\usepackage[a4paper,margin=2.2cm]{geometry}
\usepackage[T1]{fontenc}
\usepackage{mathptmx}
\usepackage{amsmath,bm}
\usepackage{graphicx}
\usepackage[super,sort&compress,comma]{natbib}
\usepackage[font=small,labelfont=bf,labelsep=space,justification=justified]{caption}
\usepackage{xcolor}
\usepackage[colorlinks=true,linkcolor=blue!50!black,citecolor=blue!50!black,urlcolor=blue!50!black]{hyperref}
\usepackage{authblk}
\usepackage{microtype}
\usepackage{siunitx}
\usepackage{booktabs}

\title{\textbf{Bias-free superconducting microwave nonreciprocity\\ from a passive nonlinear resonator}}
\author[1]{Brendan MacKay}
\author[1]{Abdul Mohamed}
\author[2,3]{Andrea Al\`u}
\author[1]{Shabir Barzanjeh\thanks{Corresponding author. E-mail \href{mailto:shabir.barzanjeh@ucalgary.ca}{shabir.barzanjeh@ucalgary.ca}}}
\affil[1]{Department of Physics and Astronomy, University of Calgary, Calgary, Alberta, Canada}
\affil[2]{Photonics Initiative, Advanced Science Research Center, City University of New York, New York, NY, USA}
\affil[3]{Physics Program, Graduate Center, City University of New York, New York, NY, USA}
\date{}

\begin{document}
\maketitle

\begin{quote}
\small\noindent
Routing microwave tones in a nonreciprocal fashion in a superconducting chip usually requires an external bias, commonly in the form of static magnetic fields, pump tones or temporal modulation. Here we demonstrate microwave nonreciprocity in a bias-free, passive platform, a single nonlinear superconducting resonator patterned in one film without Josephson junctions. Unequal coupling with the two feeding ports makes its nonlinear resonance shift direction-dependent, and its sharp Fano lineshape converts this asymmetric response into a large transmission contrast when excited from opposite directions. Our niobium nitride (NbN) device reaches a directional contrast of 53.0\,dB at an insertion loss of \SI{4.6}{dB}, and it exceeds \SI{20}{dB} over 9.7\,MHz. Time-resolved measurements show that the device routes pulses from \SI{2}{\micro s} to \SI{1}{ms} with more than \SI{30}{dB} of contrast and switches within a few microseconds. Made with the same technology as superconducting amplifiers, such bias-free nonreciprocal components could simplify the on-chip routing of strong pump and control tones in quantum processors.
\end{quote}
Superconducting quantum processors require components that transmit microwave signals in one direction only\cite{Barzanjeh2026}. Today this is usually done with ferrite circulators and isolators, which protect qubits and readout resonators from the noise of the amplifier chain and route signals between components\cite{Krantz2019,Ranzani2019}. These devices are bulky, rely on permanent magnets and add loss to every channel\cite{Pozar}, a serious cost as processors grow to thousands of qubits. Some signals, such as the pump tones of kinetic-inductance amplifiers, are themselves strong, and directionally routing them on chip without magnets would already simplify the cryogenic wiring.

Lorentz reciprocity states that any linear, time-invariant and non-magnetic device transmits equally in opposite directions\cite{Jalas2013,Caloz2018}. On-chip devices therefore break reciprocity with an external bias, such as parametric pumps\cite{Kamal2011,Abdo2013,Sliwa2015,Lecocq2017}, frequency conversion with delay\cite{Kerckhoff2015,Chapman2017,Rosenthal2017}, parametric modulation of coupled-resonator loops\cite{Estep2014}, optomechanical and electromechanical interference\cite{Barzanjeh2017,Bernier2017,Peterson2017,Ruesink2018} or quantum Hall gyrators\cite{Viola2014,Mahoney2017}. These devices isolate even weak signals, but they need an auxiliary pump, modulation or magnetic field, with the associated wiring and heat load.

Nonlinearity offers a way around reciprocity without an external bias\cite{Fan2012,DelBino2018,Cotrufo2024,RosarioHamann2018}. If a nonlinear resonator is coupled asymmetrically to two ports, signals from the two ports shift its resonance by different amounts, and a Fano resonance, whose transmission rises from a zero to a maximum within about one linewidth, turns this frequency offset into a large transmission contrast\cite{Fan2003,Limonov2017}. With a single nonlinear resonance, such devices must trade forward transmission against transmission asymmetry\cite{Sounas2017PRL,Sounas2018PRB,Sounas2018NE,Yang2020} and work only within a range of input powers and for signals from one port at a time\cite{Shi2015,Cotrufo2021a,Cotrufo2021b}. Despite these limitations, this bias-free and passive route has been demonstrated in photonic structures\cite{Fan2012} and in a free-space optical metasurface\cite{Cotrufo2024}; a flux-biased variant has been realized as a superconducting diode\cite{Upadhyay2024}, and the effect has been predicted for high-kinetic-inductance resonators\cite{Gaiser2026}. 

\begin{figure}[!t]
\centering
\includegraphics[width=\textwidth]{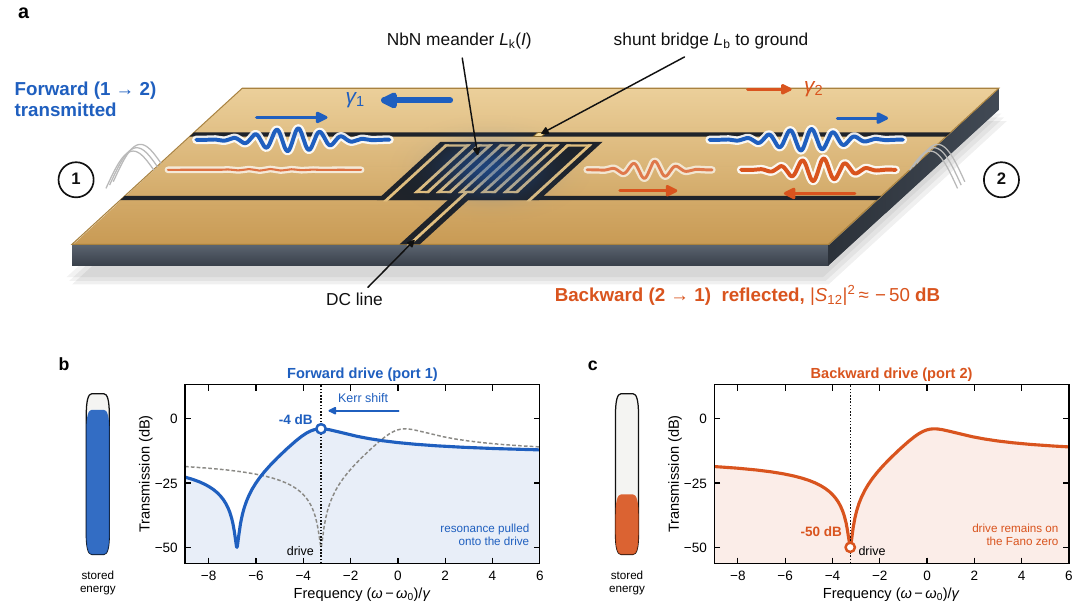}
\caption{\textbf{A single-resonator nonreciprocal microwave device.} \textbf{a}, Illustration of the device. A coplanar feed line on a NbN film has a wide centre conductor between ground planes A and B. A kinetic-inductance resonator, a serpentine NbN meander whose kinetic inductance $L_{\rm k}(I)$ grows with the current $I$, sits in a pocket of the centre conductor and is connected to ground plane B. Next to the meander the centre conductor narrows to a strip, at whose centre a narrow NbN bridge $L_{\rm b}$ shorts the line to ground plane A. The strip and the bridge together form a partially reflecting background. The resonator decays into the two halves of the line at rates $\gamma_1>\gamma_2$. A forward wave from port~1 is transmitted, whereas a backward wave of the same power from port~2 is reflected and its transmission is suppressed to about \SI{-50}{dB}. \textbf{b},\textbf{c}, Transmission seen by a forward (\textbf{b}) and a backward (\textbf{c}) drive of equal power, computed with the measured linear parameters and a pure Kerr shift, as an idealized illustration. The forward drive stores more energy (gauges) and pulls the Fano response (solid) away from its linear position (dashed), so that the drive sits near the Fano maximum. The backward response stays essentially linear and the drive sits on its zero.}
\label{fig:concept}
\end{figure}

Thin superconducting films with a large kinetic inductance, such as niobium nitride (NbN), niobium titanium nitride (NbTiN) and titanium nitride (TiN), have a fast, low-loss Kerr nonlinearity at gigahertz frequencies\cite{Zmuidzinas2012,Niepce2019,Davoudi2025}, which is the basis of kinetic-inductance amplifiers\cite{Eom2012,Malnou2021,Mohamed2024}. Here we show that a single NbN resonator is enough to build a microwave diode, by which we mean a two-port element whose transmission depends on the direction of the wave above a threshold power\cite{Fan2012,Upadhyay2024}, not a rectifier. The kinetic-inductance nonlinearity of the resonator, combined with Fano interference between its resonant path and a broad background path, makes the transmission depend on direction. The signal itself supplies the energy that breaks reciprocity, so no auxiliary tone is needed, but the signal must be strong enough to shift the resonance. Earlier superconducting realizations rely on Josephson junctions or an external bias, whereas this device consists of a single resonator in a single film, defined in one lithography step (Supplementary Information). A coupled-mode theory constrained by energy conservation and time-reversal symmetry gives a closed-form contrast ceiling, a power-rescaling symmetry and design rules, and shows that a weak coherent loss channel makes the Fano zero deeper than a two-port resonator allows. Finally, time-resolved measurements, which had not been performed on passive nonlinear diodes in superconducting circuits, show that the device routes pulses from microseconds to milliseconds and that a delayed part of its nonlinearity sets the switching time.

\begin{figure}[!t]
\centering
\includegraphics[width=\textwidth]{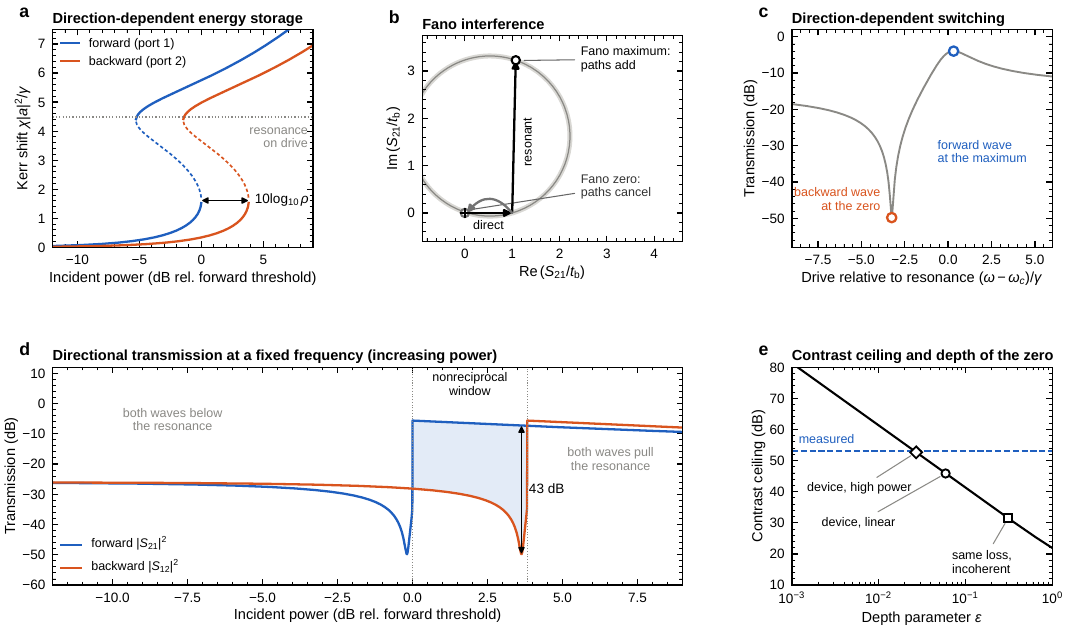}
\caption{\textbf{Mechanism of the directional contrast.} Panels \textbf{a}--\textbf{d} are idealized predictions with the measured linear parameters and a pure Kerr shift, at the drive detuning of the measured contrast maximum, $\omega-\omega_0=-4.5\gamma$, where $\omega$ is the drive frequency, $\omega_0$ the low-power resonance frequency and $\gamma$ the half-linewidth. \textbf{a}, Direction-dependent energy storage. Kerr shift of the resonance versus incident power for a drive at port~1 (blue) and port~2 (orange). Solid lines are stable and dashed lines unstable states. The resonator stores $\rho=\gamma_1/\gamma_2$ times more energy for a drive from port~1, so its response is that of port~2 shifted by $10\log_{10}\rho$ in power. \textbf{b}, Fano interference. Transmission in the complex plane, normalized to the direct path (black arrow along the real axis). The wave re-radiated by the resonator traces the grey circle as the frequency changes, nearly cancelling the direct path at the Fano zero (grey arrow) and adding to it at the maximum (black arrow). \textbf{c}, Direction-dependent switching. Linear Fano profile versus the detuning of the drive from the Kerr-shifted resonance that each wave experiences. The forward wave pulls the resonance onto itself and sits near the maximum (blue), whereas the backward wave barely shifts the resonance and stays on the zero (orange). \textbf{d}, Resulting transmission in both directions at a fixed drive frequency as the power increases. Inside the nonreciprocal window, of width $10\log_{10}\rho$, the forward wave has switched to the high-transmission state while the backward wave remains near its zero. The measured contrast exceeds this prediction, and the measured power offset (2.43\,dB) is smaller than $10\log_{10}\rho$. \textbf{e}, Inferred contrast ceiling of Eq.~\eqref{eq:ceiling} versus the depth parameter $\varepsilon$ at the measured Fano visibility $c=3.25$ (both defined in Eq.~\eqref{eq:fano}). Symbols mark the device in the linear regime, the effective value at the highest drive, and the value expected if the same loss were entirely incoherent. The dashed line is the largest measured contrast.}
\label{fig:mechanism}
\end{figure}

\section*{Operating principle}
In the device of Fig.~\ref{fig:concept}a, a coplanar feed line connects ports~1 and 2. Close to the resonator, its centre conductor narrows into a short strip that is grounded at its centre by a narrow NbN bridge. The strip and the bridge act as a partial reflector, and the small fraction they transmit forms the background of the Fano resonance. The strip alone already gives a Fano response, while the bridge increases the reflection and, according to electromagnetic simulations, keeps the background nearly flat across the resonance.

The resonator is a serpentine NbN meander, whose kinetic inductance, and hence resonance frequency, depends on the current it carries. The meander sits beside the strip, slightly offset towards port~1, and couples to the feed line both through the strip and through the adjacent coplanar sections. Its stored energy therefore decays faster into port~1, at a rate $\gamma_1$, than into port~2, at a rate $\gamma_2$, and we call $\rho=\gamma_1/\gamma_2$ the coupling asymmetry. Time-reversal symmetry permits such unequal coupling only on a reflecting line, since for a two-port line it limits the power transmission without the resonator, $|t_{\rm b}|^2$, to $|t_{\rm b}|^2\leq4\rho/(1+\rho)^2$ (Supplementary Information). The diode relies on three effects, illustrated in Fig.~\ref{fig:concept}b,c and analysed in Fig.~\ref{fig:mechanism}.

The first effect is direction-dependent energy storage. A port through which the resonator decays quickly also feeds it efficiently, again as a consequence of time-reversal symmetry. For a drive of frequency $\omega$ and power $P_j$ at port $j$, the energy stored in the resonator is $|a|^2=2\gamma_jP_j/[(\omega-\omega_c)^2+\gamma^2]$, where $a$ is the resonator amplitude, $\omega_c$ the resonance frequency and $\gamma=\gamma_1+\gamma_2+\gamma_{\rm i}+\gamma_{\rm p}$ the half-linewidth, set by the decay into the two ports, internal loss and a channel introduced below (Methods). The same wave therefore stores $\rho$ times more energy when it arrives from port~1 than from port~2. The kinetic inductance converts this energy into a downward shift of the resonance, $\omega_c=\omega_0-\chi|a|^2$, where $\omega_0$ is the low-power resonance frequency and $\chi>0$ the Kerr coefficient. A drive below the resonance thus pulls the resonance towards itself and stores even more energy. More than $\sqrt3$ half-linewidths below the resonance this feedback makes the response bistable, and above a threshold power the resonance jumps onto the drive\cite{Siddiqi2005}. The stored energy depends on the drive only through $\gamma_jP_j$, so a forward wave reaches this threshold at a power $10\log_{10}\rho$ lower than a backward wave (Fig.~\ref{fig:mechanism}a).

The second effect is Fano interference. The wave that reaches the far port is the sum of a direct part, transmitted past the strip and the bridge, and a resonant part, re-emitted by the resonator. As the drive frequency crosses the resonance, the resonant part rotates in phase and the total traces a circle in the complex plane (Fig.~\ref{fig:mechanism}b). The two parts cancel close to the origin, giving a deep minimum, the Fano zero, and add on the opposite side, giving the transmission maximum.

\begin{figure}[!t]
\centering
\includegraphics[width=\textwidth]{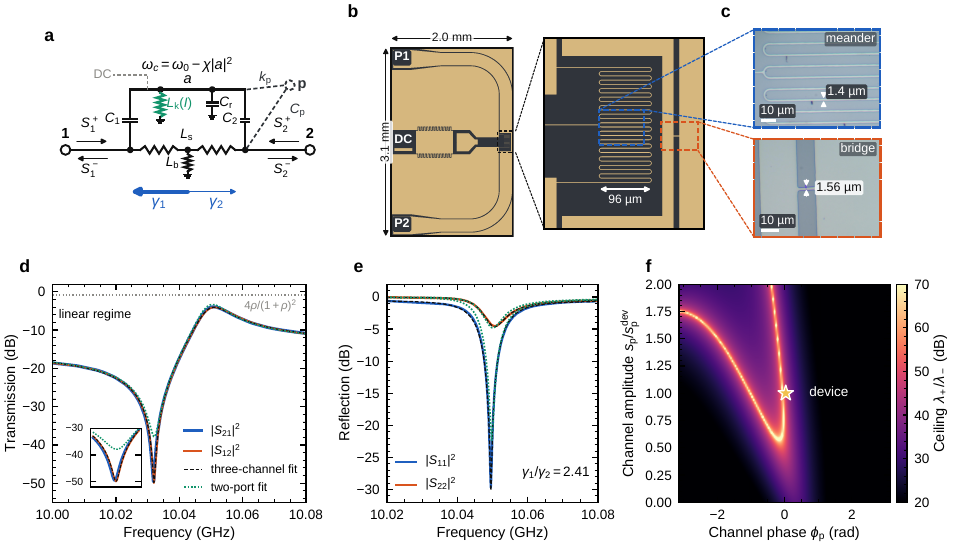}
\caption{\textbf{Device and linear response.} \textbf{a}, Circuit model. A resonator formed by the nonlinear kinetic inductance $L_{\rm k}(I)$ and its capacitance $C_{\rm r}$ couples to the port-1 and port-2 sides of the feed line through the coupling capacitances $C_1>C_2$. The narrow strip $L_{\rm s}$ and the shunt bridge $L_{\rm b}$ at its centre form the partially reflecting background, and the resonator decays into the two ports at rates $\gamma_1>\gamma_2$. Dashed, undetected coherent channel p, fed by the resonator ($k_{\rm p}$) and by the direct path ($C_{\rm p}$, representing $C_{1{\rm p}}$ and $C_{2{\rm p}}$; Supplementary Information). $S_j^\pm$ are the incoming and outgoing waves and $P_j$ the incident power at port $j$. \textbf{b}, Layout of the measured device (dark, etched; light, NbN), a \SI{2.0}{mm}\,$\times$\,\SI{3.1}{mm} region containing the feed line between ports 1 and 2 (P1, P2) and the filtered direct-current (DC) line, rotated by \SI{180}{\degree}. The enlarged view shows the resonator pocket, the meander and the slot crossed by the shunt bridge. \textbf{c}, Unprocessed optical micrographs of the meander (blue) and the shunt bridge (orange) marked in \textbf{b}, with widths measured by the microscope software (thin lines). NbN appears bright and the etched substrate grey. \textbf{d}, Calibrated transmission in both directions in the linear regime (\SI{-30}{dBm} source power) with the three-channel coupled-mode fit (dashed) and the best fit restricted to the two signal ports (dotted green). The inset enlarges the transmission zero, which the two-port model misses by \SI{12}{dB}. The dotted line marks the lossless two-port bound $4\rho/(1+\rho)^2$. \textbf{e}, Calibrated reflection at the two ports with the same fits. The deep dip at port~1 and the shallow dip at port~2 give $\rho=\gamma_1/\gamma_2=2.41$. \textbf{f}, Model calculation of the contrast ceiling $\lambda_+/\lambda_-$ from the fitted parameters while the two coupled-mode parameters that control the third channel, its coupling strength $s_{\rm p}$ and phase $\phi_{\rm p}$ (Methods), are varied with all others fixed. This also changes the signal-port couplings and the background moderately (Supplementary Information). Bright lines mark perfect Fano zeros ($\varepsilon=0$). The star marks the device.}
\label{fig:device}
\end{figure}

The third effect is the combination of the two. With the drive placed just below the Fano zero, a backward wave stores little energy, barely shifts the resonance and remains on the zero. A forward wave of the same power pulls the resonance onto itself and moves to the opposite side of the circle, close to the maximum (Figs.~\ref{fig:concept}b,c and \ref{fig:mechanism}c,d). This does not contradict Lorentz reciprocity, which applies only to linear systems. The fixed-parameter calculation of Fig.~\ref{fig:mechanism} yields \SI{43}{dB}, whereas the measured contrast reaches 53.0\,dB, as discussed below.

The largest possible contrast follows from the Fano form of the forward transmission $S_{21}$, the wave amplitude leaving port~2 for a unit wave entering port~1\cite{Fan2003,Limonov2017},
\begin{equation}
|S_{21}|^2=|t_{\rm b}|^2\,\frac{\varepsilon^2+(y+c)^2}{1+y^2},\qquad y=\frac{\omega-\omega_c}{\gamma}, \label{eq:fano}
\end{equation}
where $y$ is the dimensionless detuning and $t_{\rm b}$ the complex amplitude of the direct, non-resonant transmission. The Fano visibility $c$ places the zero at $y=-c$, and the depth parameter $\varepsilon$ sets how close the circle comes to the origin. For small $\varepsilon$ the maximum lies near $y=1/c$, so for $|c|>1$, as here, the resonance lies much closer to the maximum than to the zero. The largest and smallest values of $|S_{21}|^2$ are $|t_{\rm b}|^2\lambda_+$ and $|t_{\rm b}|^2\lambda_-$, with $\lambda_\pm=\tfrac12\bigl[\sigma\pm\sqrt{\sigma^2-4\varepsilon^2}\bigr]$ and $\sigma=1+c^2+\varepsilon^2$ (Methods). Placing the forward wave at the maximum and the backward wave at the zero gives the largest contrast $\mathcal{C}=|S_{21}|^2/|S_{12}|^2$, where $S_{12}$ is the backward transmission,
\begin{equation}
\mathcal{C}_{\max}=\frac{\lambda_+}{\lambda_-}\simeq\frac{(1+c^2)^2}{\varepsilon^2}, \label{eq:ceiling}
\end{equation}
where the approximation holds for $|\varepsilon|\ll\sqrt{1+c^2}$. This ceiling depends only on the low-power response and grows as $1/\varepsilon^2$ (Fig.~\ref{fig:mechanism}e). For a resonator coupled only to the two signal ports, the depth is set by the internal loss, $\varepsilon=\gamma_{\rm i}/\gamma$, where $\gamma_{\rm i}$ is the internal loss rate (Methods), so any internal loss makes the zero shallower and lowers the ceiling. The next section shows that a weak additional channel removes this limit.


\section*{Device and linear response}
Figure~\ref{fig:device}a shows the circuit model of the device, which is patterned in a \SI{10}{nm} NbN film on silicon (Fig.~\ref{fig:device}b,c and Methods). The strip, the bridge and the adjacent coplanar sections are symmetric about the bridge, and the meander, about \SI{0.1}{mm} by \SI{0.2}{mm}, is shifted by \SI{21.5}{\micro m} towards port~1. A layout model predicts $\rho\approx2.2$ for this shift and $\rho\approx1.1$ for a meander centred on the bridge (Supplementary Information). A filtered DC port (Fig.~\ref{fig:device}b) allows the operating frequency to be tuned with a direct current (Methods).

At low power both transmissions show the same Fano profile, with a zero of -50\,dB at \SI{10.032}{GHz} and a maximum of -3.9\,dB at \SI{10.050}{GHz}. They agree to within a median of 0.10\,dB, as reciprocity requires (Fig.~\ref{fig:device}d). Reflection differs between the ports, with a dip of \SI{30}{dB} at port~1, close to critical coupling, but only \SI{4.6}{dB} at port~2 (Fig.~\ref{fig:device}e).

\begin{figure}[!t]
\centering
\includegraphics[width=\textwidth]{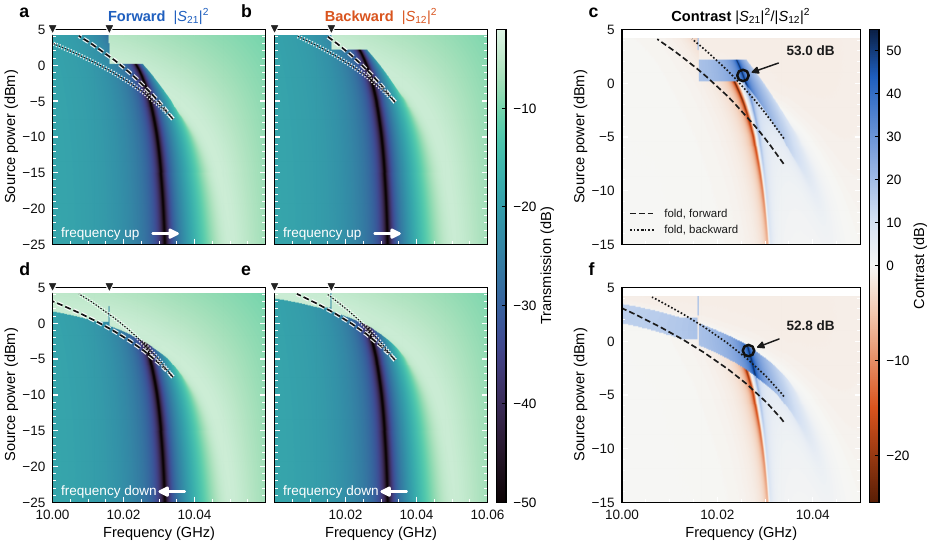}
\caption{\textbf{Directional transmission.} \textbf{a},\textbf{b}, Calibrated forward ($|S_{21}|^2$) and backward ($|S_{12}|^2$) transmission versus frequency and vector network analyser (VNA) source power for upward frequency sweeps. \textbf{d},\textbf{e}, Same for downward sweeps. Dashed lines mark the end of the branch that the model predicts for each sweep, and dotted lines the opposite edge of the bistable band. Source powers are corrected for the attenuator step of the source, which causes the horizontal discontinuity, and triangles mark two frequencies where source transients induce switching (Methods). \textbf{c},\textbf{f}, Contrast $|S_{21}|^2/|S_{12}|^2$ with the predicted switching edges of the two directions (dashed, forward; dotted, backward), on a diverging scale centred at \SI{0}{dB}, with blue for forward-favoured and orange for backward-favoured transmission. Circles mark the maxima.}
\label{fig:maps}
\end{figure}

These four traces over-determine the internal loss and therefore test the model stringently. In a resonator coupled only to the two signal ports, energy conservation and time-reversal symmetry fix the relative phase of the direct and resonant paths, so the internal loss alone sets the depths of the transmission zero and of both reflection dips (Methods). The measured depths are inconsistent with a single loss. The zero corresponds to $\varepsilon\approx0.06$, whereas the reflection dips require $\varepsilon\approx0.26$, and the best fit restricted to the two signal ports misses the zero by \SI{12}{dB} (Fig.~\ref{fig:device}d, inset). We therefore asked which additional element could reconcile the three depths while keeping the circuit passive and reciprocal. An effective coherent channel that is not connected to the measurement lines does so. Coupled to both the resonator and the direct path, it changes the phase relation that energy conservation and time-reversal symmetry impose on the two signal paths, and with it the interference at the zero. With this channel the model reproduces the magnitude and phase of all four traces with a 6.4 times smaller residual (Fig.~\ref{fig:device}d,e) and predicts each reflection trace withheld from the fit, which the two-port model misses by 8 and \SI{23}{dB} (Methods). A layout model points to a likely origin, conversion of the coplanar wave into an undetected slot-line mode at the one-sided bridge. Other origins cannot be ruled out, but our conclusions rely only on the existence of the channel, and the bridge explanation makes a testable prediction (see Design rules). The fit gives $\gamma_1/2\pi=2.60$\,MHz, $\gamma_2/2\pi=1.08$\,MHz, a leakage rate into the extra channel $\gamma_{\rm p}/2\pi=0.29$\,MHz and an internal loss rate $\gamma_{\rm i}/2\pi=1.38$\,MHz, so that $\rho=2.41$ and the full linewidth is 10.7\,MHz.

The extra channel takes only -11.0\,dB of the power incident from port~1 and about \SI{5}{\percent} of the energy that the resonator loses, yet it makes the depth parameter five times smaller, from 0.31, the value for the same loss without interference, to $\varepsilon=0.059$. The modified phase relation lets the direct and the resonant wave cancel again at the zero, even though the internal loss weakens the resonant wave. As a result, the linear ceiling rises to 45.8\,dB, 14\,dB above the value without the channel (Fig.~\ref{fig:mechanism}e). Varying the coupling strength $s_{\rm p}$ and phase $\phi_{\rm p}$ of the channel in the model (Methods) gives lines of perfect zeros, and the device lies only \SI{0.11}{rad} in $\phi_{\rm p}$ from one of them (Fig.~\ref{fig:device}f).

\begin{figure}[!t]
\centering
\includegraphics[width=\textwidth]{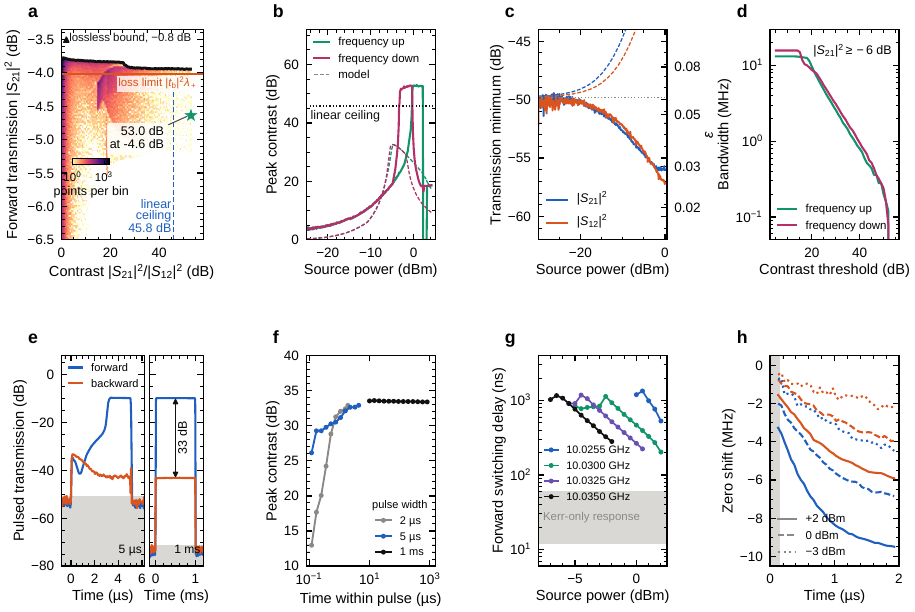}
\caption{\textbf{Performance and time-domain operation.} \textbf{a}, Operating envelope. All measured points of the maps (both sweep directions) in the plane of contrast and forward transmission, coloured by the number of points per bin (logarithmic scale). The black line is the highest forward transmission reached at each contrast. The orange line is the loss limit of the linear response, $|t_{\rm b}|^2\lambda_+$, and the dashed blue line the contrast ceiling of the linear response. The lossless two-port bound $4\rho/(1+\rho)^2$ lies above the plotted range (arrow). The star marks the largest measured contrast. \textbf{b}, Peak contrast versus source power for both sweep directions (solid, data; dashed, model). The dotted line is the linear ceiling. \textbf{c}, Depth of the transmission minimum versus source power for both directions (solid, data; dashed, model), with the corresponding depth parameter on the right axis. At high power this is an effective value inferred from the measured minimum. \textbf{d}, Largest bandwidth at a single source power over which the contrast exceeds a given value while the forward transmission stays above \SI{-6}{dB}. Panels \textbf{a}--\textbf{d} were measured with continuous tones and the DC port unbonded. \textbf{e}--\textbf{h}, Pulsed measurements in a later configuration with the DC port bonded (Methods). \textbf{e}, Forward (blue) and backward (orange) transmission for a \SI{5}{\micro s} (left) and a \SI{1}{ms} (right) pulse at \SI{10.0245}{GHz} and \SI{+0.5}{dBm}, relative to the bypass. Grey bands mark the detection floor, which is lower for the \SI{1}{ms} pulse measured with a narrower bandwidth. \textbf{f}, Largest contrast over the measured grid in successive time windows within pulses of three widths. \textbf{g},\textbf{h}, \SI{2}{\micro s} pulses. \textbf{g}, Forward switching delay versus source power at four frequencies. The grey band is the range of delays predicted by a model with only the Kerr shift. \textbf{h}, Shift of the transmission zero during the pulse for forward (blue) and backward (orange) drive at three source powers, relative to its low-power position. The grey band marks the first \SI{150}{ns}.}
\label{fig:fom}
\end{figure}


\section*{Directional transmission}
Figure~\ref{fig:maps} shows the calibrated transmission in both directions and their ratio for upward (a--c) and downward (d--f) frequency sweeps. As the power increases, the resonance moves to lower frequency and becomes bistable\cite{Abdo2006,Swenson2013}, with a sawtooth in upward sweeps, where the resonance jumps onto the drive, and dragging in downward sweeps. All these features appear at lower power in the forward direction. The contrast (Fig.~\ref{fig:maps}c,f) is therefore concentrated in a narrow ridge along the switching edges, where the two directions are on different branches (Fig.~\ref{fig:mechanism}d). A minimal model that adds only a Kerr shift and a nonlinear loss to the low-power response, with two fitted parameters (Methods), reproduces the onset of bistability and the switching edges to within a few megahertz (lines in Fig.~\ref{fig:maps}), with small deviations from noise-activated escape\cite{Siddiqi2005} (Supplementary Information).

The largest contrast is 53.0\,dB, at \SI{10.0254}{GHz} and +0.7\,dBm source power in an upward sweep, with -4.6\,dB forward and -57.6\,dB backward transmission. Downward sweeps reach 52.8\,dB, 651 points exceed \SI{40}{dB}, and equal chip-port powers still give 52.3--52.8\,dB, with an uncertainty of roughly $\pm\SI{1}{dB}$ (Supplementary Information).

The data also obey a symmetry that tests the mechanism without a model. A single-mode resonator responds only to the energy it stores, not to the port it came from, so that $S_{21}(\omega,P)=S_{12}(\omega,\rho P)$, where $P$ is the incident power (Supplementary Information). Shifting the backward maps along the power axis indeed makes them fall onto the forward maps, including the bistable jumps. The best shift, 2.43\,dB, reduces the mean difference between the maps by a factor of 7.2. It is 1.4\,dB smaller than $10\log_{10}\rho=3.8$\,dB (Supplementary Information), so the collapse yields an effective drive ratio rather than an independent measure of $\rho$, and we use it to fix this ratio in the nonlinear model. The nonreciprocity is therefore a power offset between the two directions.


\section*{Performance}
Over most of the contrast range the best forward transmission stays within \SI{0.25}{dB} of the loss limit set by the low-power response, $|t_{\rm b}|^2\lambda_+=-4.0$\,dB (Fig.~\ref{fig:fom}a). This limit is set by the internal loss and the leakage, about \SI{30}{\percent} of the linewidth, and lies \SI{3.2}{dB} below the lossless two-port bound $4\rho/(1+\rho)^2$\cite{Sounas2017PRL}.

The measured contrast exceeds the linear ceiling of 45.8\,dB because the transmission minimum becomes deeper at high power (Fig.~\ref{fig:fom}b,c). The minimum stays at -50\,dB up to about \SI{-25}{dBm} and reaches \SI{-56}{dB} near \SI{0}{dBm}, where the backward wave sits. In terms of Eq.~\eqref{eq:ceiling}, the effective depth parameter falls from 0.059 to 0.026, which corresponds to a ceiling of about \SI{53}{dB} (Fig.~\ref{fig:mechanism}e). The minimal model, whose nonlinear loss makes the zero shallower, reaches only 33\,dB (Supplementary Information). It also underestimates the high-energy branch (\SI{-8}{dB} against \SI{-4}{dB} measured), so the highest contrast comes from a deepening of the zero that the model does not capture. A phenomenological extension follows this deepening up to about \SI{-10}{dBm} (Supplementary Information), and a saturating internal loss\cite{Muller2019} or a drift towards a line of perfect zeros (Fig.~\ref{fig:device}f) could explain it, but we do not assign a mechanism.

\begin{figure}[!t]
\centering
\includegraphics[width=\textwidth]{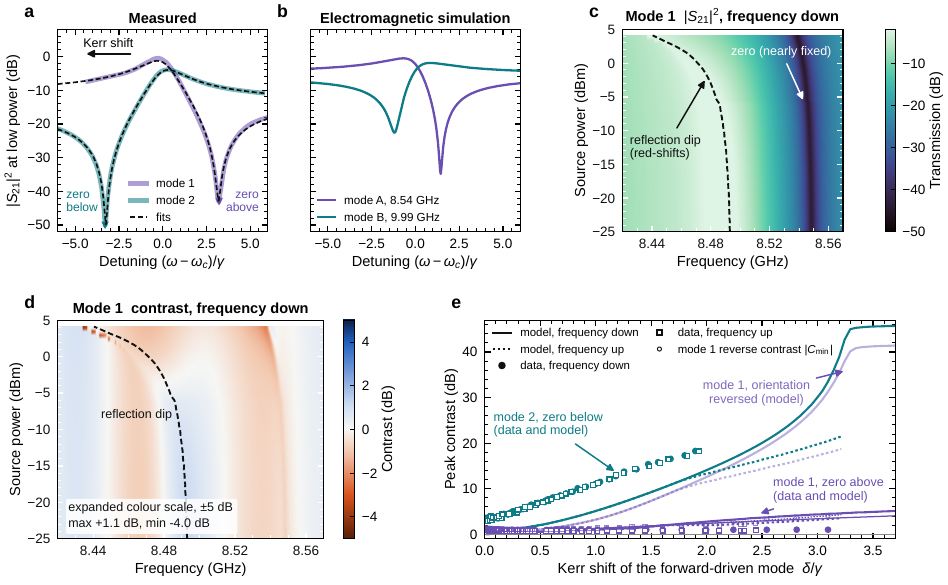}
\caption{\textbf{Orientation of the Fano profile as a design rule.} \textbf{a}, Calibrated low-power transmission of the two modes of the resonator versus detuning in units of the half-linewidth. Dashed black lines are the coupled-mode fits. The fit places the maximum of mode~1 at \SI{-1.3}{dB}, set by its internal loss and leakage, and the measured trace lies 0.4--\SI{0.8}{dB} higher, within the calibration uncertainty of this mode (Supplementary Information). Mode~1 (8.493\,GHz) has its zero above the resonance, mode~2 (10.0494\,GHz) below it (triangles). The arrow marks the direction of the Kerr shift. \textbf{b}, The same for the two lowest modes of an electromagnetic simulation of the measured layout at a sheet inductance of \SI{58}{pH/sq}, the value that best reproduces the measured mode frequencies. \textbf{c}, Forward transmission map of mode~1 for a downward frequency sweep. The dashed line is the frequency of the port-1 reflection dip, which follows the Kerr shift of the resonance. The transmission zero stays nearly fixed while the resonance shifts by three half-linewidths. \textbf{d}, Contrast map of mode~1 on an expanded colour scale of $\pm\SI{5}{dB}$, with the reflection dip (dashed). The contrast never exceeds a few decibels. \textbf{e}, Peak contrast versus the largest measured resonance shift $\delta$ of the forward-driven mode, in units of its half-linewidth $\gamma$, for both modes and both sweep directions. Symbols, data; lines, single-mode model with the linear parameters of each mode. The light line is a hypothetical model of mode~1 with the sign of its visibility reversed. Mode~2 data are shown up to the onset of bistability, beyond which the reflection dip no longer tracks the resonance.}
\label{fig:parity}
\end{figure}

With forward transmission above \SI{-6}{dB} (Fig.~\ref{fig:fom}d), the contrast exceeds \SI{20}{dB} over 9.7\,MHz and \SI{30}{dB}, \SI{40}{dB} and \SI{50}{dB} over 3.1, 1.0 and 0.24\,MHz. The peak contrast stays above \SI{20}{dB} over a \SI{6.9}{dB} window of source power. At a fixed frequency the window is 2.1--\SI{2.4}{dB}, close to the power offset of 2.43\,dB between the two directions. At the contrast maximum the chip receives about -42\,dBm (\SI{60}{nW}) and the resonator holds a few times $10^{7}$ photons (Methods).

\section*{Time-domain operation}
As the nonlinear response does not obey superposition, we also sent pulses of \SI{2}{\micro s}, \SI{5}{\micro s} and \SI{1}{ms} to either port and recorded the transmitted power in real time (Methods). With the DC port bonded and held at zero current, the internal loss rises from 1.38 to about \SI{1.5}{MHz} and the steady-state contrast measured with continuous tones reaches about \SI{35}{dB} instead of 53.0\,dB. By Eq.~\eqref{eq:ceiling}, the added loss alone lowers the linear ceiling by only about \SI{2}{dB}, so the bond probably also alters the coherent channel (Fig.~\ref{fig:device}f). At \SI{10.0245}{GHz} and \SI{+0.5}{dBm} a \SI{5}{\micro s} forward pulse ends at \SI{-9.6}{dB} and the backward pulse at \SI{-42.6}{dB}, a contrast of \SI{33}{dB} within a single pulse (Fig.~\ref{fig:fom}e). The \SI{2}{\micro s} pulses give the same value. The contrast builds up within about \SI{2}{\micro s} and then stays between 33.3 and \SI{33.6}{dB} throughout a \SI{1}{ms} pulse, while the forward level drifts by a median of only \SI{0.05}{dB} (Fig.~\ref{fig:fom}e,f). The narrower detection bandwidth of these long pulses places the backward level \SI{30}{dB} above the floor, so the contrast is set by the device and not by the instrument. The power-rescaling collapse holds at every time within the pulse, with a shift of about \SI{3.3}{dB} (Supplementary Information).

The forward wave starts on the zero and switches to the high-transmission state after 0.2--\SI{1.6}{\micro s}, a delay that shortens as the power increases and reaches \SI{3.7}{\micro s} close to the threshold in \SI{5}{\micro s} pulses (Fig.~\ref{fig:fom}g). A model with only the Kerr shift and the measured rates predicts 10--\SI{60}{ns}. Successive pulses are almost identical, so the slow switching is deterministic. Tracking the transmission zero (Fig.~\ref{fig:fom}h) shows that it moves by only 15--\SI{35}{\percent} of its final shift within the first \SI{150}{ns}, and that the rest builds up with a time constant of 0.6--\SI{3}{\micro s}. The resonance shift therefore has a prompt part, compatible with the intrinsic kinetic-inductance Kerr nonlinearity within our \SI{65}{ns} resolution, and a larger delayed part. The stationary shift of the continuous-wave maps, and the Kerr coefficient fitted to them, are therefore effective quantities that contain both parts. The collapse holds for the delayed part as well, so it is driven by the energy absorbed in the resonator, pointing to local heating of the film or nonequilibrium quasiparticles rather than to the measurement lines. This delayed part limits the switching speed and sets a trade-off between pulse length and power, as contrast above \SI{20}{dB} requires \SI{+5}{dBm} within the first \SI{250}{ns} but only \SI{-3.5}{dBm} once the response has settled (Supplementary Information).

\section*{Design rules}

Our analysis gives three rules for designing Kerr--Fano diodes. The first rule concerns the orientation of the Fano profile. The Kerr shift lowers the resonance, so a drive at a zero below the resonance pulls the resonance towards itself and the forward response switches from the zero to the maximum (Fig.~\ref{fig:mechanism}c). If the zero lies above the resonance, the drive pushes the resonance away and the response changes only gradually.

The lower mode of the same resonator tests this rule. This mode, at 8.493\,GHz, also couples more strongly to port~1 ($\rho=1.40$), but its zero lies above the resonance in both the measurement and an electromagnetic simulation (Fig.~\ref{fig:parity}a,b). Under strong drive it shifts by 3.1 half-linewidths while its zero stays almost fixed (Fig.~\ref{fig:parity}c), and its contrast stays between $-4.0$ and 1.1\,dB (Fig.~\ref{fig:parity}d). At the same Kerr shift, mode~2 reaches about \SI{20}{dB}. The modes also differ in coupling ratio and linewidth, but reversing only the orientation of mode~1 in the model raises its contrast to about 41\,dB (Fig.~\ref{fig:parity}e). The orientation alternates from one meander mode to the next (Supplementary Information), so the right mode can be chosen at the design stage.

The second rule concerns the depth of the zero. A coherent channel can make $\varepsilon$ much smaller than the lost fraction of the linewidth, and thus becomes a design parameter. If it comes from the one-sided bridge, a second bridge across the other slot should suppress it and make the zero shallower, with $\varepsilon$ close to $\gamma_{\rm i}/\gamma$, which would directly test this explanation (Methods). Conversely, the model indicates that an engineered auxiliary port or resonator tuned onto a line of perfect zeros could raise the contrast well above \SI{60}{dB}.

The third rule concerns the insertion loss. The loss limit $|t_{\rm b}|^2\lambda_+$ does not improve with a deeper zero, and lowering it requires an internal quality factor above our 3,640. At $3\times10^4$, a value reached in NbN resonators\cite{Niepce2019}, the loss limit of the same design drops to about \SI{1.9}{dB}.

\section*{Conclusions}
We have shown that a single kinetic-inductance resonator, made in one superconducting film in one lithography step, works as a nonlinear microwave diode that needs no magnetic field, bias, auxiliary pump or modulation. It reaches 53.0\,dB of contrast at -4.6\,dB forward transmission and more than \SI{20}{dB} over 9.7\,MHz, and it routes pulses from \SI{2}{\micro s} to \SI{1}{ms} with more than \SI{30}{dB} of contrast. Its nonreciprocal element occupies about \SI{0.02}{mm^2}, and unlike other on-chip approaches it needs neither a Josephson junction nor any resource other than the signal itself (Supplementary Information).

The mechanism also sets the limits of the device. The contrast exists only within a window of input powers, and only for signals that arrive from one port at a time\cite{Shi2015,Cotrufo2021b}. A weak backward signal that arrives together with a strong forward wave sees a small-signal response that is reciprocal to leading order, so the strong wave does not block it\cite{Shi2015}. The device therefore cannot direct thermal noise\cite{Barzanjeh2018} and does not replace the ferrite isolators that protect qubits.

Made in the same film as kinetic-inductance resonators and amplifiers\cite{Eom2012,Malnou2021}, the diode can share their chip without magnets, ferrites or extra control lines. It is best viewed as a power-selective router for strong tones that arrive from one port at a time. It could route pump or control pulses on chip, passing them in one direction while blocking the same tones sent from the other side\cite{Barzanjeh2026,Ahmadi2024}. Such passive routing would reduce the ferrite hardware in large cryostats, provided the drive is kept away from nearby qubits. Reducing the slow part of the nonlinearity and measuring a weak backward probe under a strong forward pulse are the next steps.

The present operating power is similar to the pump power of kinetic-inductance travelling-wave amplifiers but far above typical readout tones and qubit drives. The threshold scales as $\gamma^3/(\gamma_j\chi)$, so a narrower linewidth and a larger Kerr shift per photon could lower it by orders of magnitude and extend the same functions to control pulses and readout tones. The DC port shifts the Fano resonance by several linewidths while preserving its orientation and the spacing of zero and maximum (Methods), so the operating frequency can be matched to a given tone in situ.

More broadly, our results show that the contrast of a nonlinear diode is limited by the depth of a Fano zero, and that this depth can be set by a weak coherent channel rather than by the quality factor alone.

\section*{Methods}
\subsection*{Device design and fabrication}
The device was designed with Sonnet \texttt{em} (v19.52). The feed line and the DC filter are defined by etched trenches in a continuous NbN ground plane. The resonator is a meandering NbN wire whose two outer arms are connected to ground. It sits in a pocket cut into the centre conductor of the feed line. Next to it, the centre conductor narrows to a strip, and a narrow NbN bridge across one slot (designed \SI{2}{\micro m} wide and \SI{10}{\micro m} long) connects this strip to ground. A DC port connects to the resonator through meander inductors and capacitive stubs that suppress radio-frequency (RF) leakage to below \SI{-40}{dB} near the resonance (Supplementary Fig.~S13). The DC port was included in the design as an input for static tuning and control of the resonator. It was not wire-bonded for the main measurements and was bonded only for the DC-current and time-domain measurements, which were taken on the same chip after bonding. The film was deposited on a high-resistivity intrinsic silicon wafer (resistivity above \SI{20}{k\ohm\,cm}), and the wafer was diced into \SI{10}{mm}\,$\times$\,\SI{10}{mm} chips. The NbN film is \SI{10}{nm} thick. Simulations of the layout reproduce the measured mode frequencies, 8.493 and 10.0494\,GHz, to within \SI{60}{MHz} for a sheet kinetic inductance of \SI{58}{pH/sq}, below the value assumed in the design (Supplementary Note~13). The device is defined in a single optical lithography step. Each chip was cleaned in acetone and isopropanol (IPA) baths for \SI{5}{min} each, dehydrated on a hotplate at \SI{100}{\celsius}, spin-coated with AZ1512 photoresist and soft-baked at \SI{100}{\celsius} for \SI{1}{min}. The pattern was exposed with a maskless laser writer (Heidelberg Instruments MLA~150) and developed in diluted AZ~400K developer. After inspection for defects under an optical microscope, the pattern was transferred into the film by reactive-ion etching in a sulfur hexafluoride (SF$_6$) plasma, and the remaining resist was stripped in acetone and IPA. Optical microscopy shows line widths of \SIrange{1.28}{1.59}{\micro m} (mean \SI{1.43}{\micro m}) in the meander and a \SI{1.56}{\micro m} wide bridge, consistent with lateral over-etching of about \SI{0.25}{\micro m} per edge. The two ground planes are connected only by rows of wire bonds that cross the feed line on both sides of the resonator, about \SI{0.3}{mm} from the slots and at a pitch of 150--\SI{200}{\micro m}. There are no air bridges.

\subsection*{Measurement setup}
The chip was mounted in a sample box made of oxygen-free high-conductivity (OFHC) copper on the mixing-chamber plate of a dilution refrigerator, which was at about \SI{10}{mK}. Two cryogenic coaxial switches (Radiall) connect an attenuated input line and an output line, which carries a cryogenic low-noise amplifier, to the two chip ports through a circulator and an isolator at each port, in two identical branches, in either transmission orientation, in a reflection configuration at either port, or through a bypass (Supplementary Fig.~S1). All scattering parameters were recorded with a VNA (Rohde \& Schwarz ZNB) with \SI{1}{kHz} intermediate-frequency (IF) bandwidth and 2001 points between \SI{10.00}{GHz} and \SI{10.08}{GHz}. The linear response was obtained by averaging many repeated traces at the lowest source power of \SI{-30}{dBm}. Transmission maps cover \SI{-30}{dBm} to \SI{+5}{dBm} nominal source power, or \SI{+4.1}{dBm} after the step correction described below, in steps of \SI{0.1}{dB}, and reflection maps \SI{-30}{dBm} to \SI{-5}{dBm} in steps of \SI{0.5}{dB}, each for upward and downward frequency sweeps at increasing power. The first point of every trace is an acquisition artefact and was discarded. The internal attenuator of the source switches at \SI{-5}{dBm} and lowers the delivered power by \SI{0.88}{dB}. We measured this step from the data, because the maps just above and just below the switching point coincide when the powers above it are lowered by 0.85--\SI{0.90}{dB} (residual \SI{0.07}{dB}) for both directions and both sweep directions. All source powers above \SI{-5}{dBm} are corrected by this measured step, and the nonlinear model contains no step. Switching events occur at the same two frequencies in all maps. The event at \SI{10.000}{GHz} coincides with an internal band boundary of the analyser source and is most likely triggered by a source transient, whereas the origin of the event at \SI{10.016}{GHz} is unclear. Neither depends on power or direction, and both are excluded from the fits (Supplementary Note~9).

\subsection*{Tuning with a direct current}
The DC port was designed as an input for static tuning and control of the resonator and was not bonded for the measurements above. With the port bonded, we measured the low-power transmission for direct currents up to \SI{400}{\micro A}. The Fano zero and the transmission maximum shift together as $\Delta f/f_0=-(I_{\rm dc}/I_2)^2$, where $f_0$ is the frequency at zero current and $I_2$ a fitted current scale of \SI{6.5}{mA} for the zero and \SI{6.7}{mA} for the maximum. At \SI{400}{\micro A} the zero moves by \SI{38}{MHz} and the maximum by \SI{36}{MHz}, and the orientation of the Fano profile and the spacing of zero and maximum are preserved, although the internal loss grows to about \SI{2.6}{MHz} and the zero becomes shallower at the largest current (Supplementary Note~16). With the port bonded, the internal loss at zero current, $\gamma_{\rm i}/2\pi\approx\SI{1.5}{MHz}$, is slightly higher than the 1.38\,MHz of the main measurements.

\subsection*{Time-domain measurements}
A microwave source (Rohde \& Schwarz SGS100A) produced \SI{2}{\micro s} pulses with a period of \SI{20}{\micro s}. A spectrum analyser (Rohde \& Schwarz FSW) in zero-span mode at the carrier frequency, triggered by the source, recorded the output power with an RMS detector and a \SI{10}{MHz} resolution bandwidth, which gives a 10--\SI{90}{\percent} rise time of about \SI{65}{ns}. Each trace has 1001 points over \SI{3}{\micro s} and is the power average of 30 pulses, for which the minimum and maximum were also stored. We measured a coarse grid from \SIrange{9.94}{10.14}{GHz} in \SI{2}{MHz} steps and \SIrange{-25}{+5}{dBm} in \SI{2}{dB} steps, and a fine grid from \SIrange{10.00}{10.04}{GHz} in \SI{0.5}{MHz} steps and \SIrange{-8}{+2}{dBm} in \SI{0.5}{dB} steps, each in both directions and through the bypass. Every trace was divided by the bypass level at the same frequency and power. Pulses of \SI{5}{\micro s} (period \SI{50}{\micro s}, \SI{10}{MHz} resolution bandwidth) and \SI{1}{ms} (period \SI{10}{ms}, \SI{100}{kHz} resolution bandwidth, rise time about \SI{7}{\micro s}) were measured from \SIrange{10.005}{10.050}{GHz} in \SI{0.5}{MHz} steps and from \SIrange{-10}{+5}{dBm} in \SI{0.5}{dB} steps. For these, the bypass was recorded at \SI{-5}{dBm} at every frequency and scaled to the other powers, and bypass traces at several powers confirmed that the through line is linear to within \SI{0.4}{dB}. The two directions differ by a route-dependent offset of $1.7\pm\SI{0.5}{dB}$ near the resonance, which we determined from the low-power traces, where the device is reciprocal, and removed from the forward data. The source powers refer to this source and are not identical to those of the network analyser. We define the switching delay as the time at which the forward trace crosses the midpoint, in decibels, between its levels at 0.10--\SI{0.20}{\micro s} and at the end of the pulse (1.75--\SI{1.95}{\micro s} for \SI{2}{\micro s} pulses and 4.0--\SI{4.75}{\micro s} for \SI{5}{\micro s} pulses). The position of the transmission zero at each time was found from the minimum over frequency in \SI{50}{ns} windows, with parabolic interpolation. The Kerr-only delays follow from the time-dependent coupled-mode equation of Supplementary Note~4 with the measured rates (Supplementary Note~17).

\subsection*{Calibration}
At thermal equilibrium every passive component at the mixing-chamber temperature $T$ radiates a noise spectral density $k_{\rm B}T_{\rm n}(T)$ into the output line, with $T_{\rm n}=(hf/2k_{\rm B})\coth(hf/2k_{\rm B}T)$\cite{Simbierowicz2021}, where $f$ is the frequency and $h$ and $k_{\rm B}$ are the Planck and Boltzmann constants. We raised $T$ from \SI{75}{mK} to \SI{1.30}{K} in ten steps and recorded the output noise at \SI{10.0493}{GHz} with a spectrum analyser (Rohde \& Schwarz FSW, \SI{100}{Hz} resolution bandwidth, root-mean-square (RMS) detector). A weighted fit of the output noise power, $N(T)=GhfB[\tfrac12\coth(hf/2k_{\rm B}T)+n_{\rm add}]$ with detection bandwidth $B$, gives an output gain $G=50.78\pm0.06$\,dB at the switch plane and an added noise of the amplifier chain of $n_{\rm add}=13.6$ quanta. We used the nominal resolution bandwidth for $B$. The effective noise bandwidth of the analyser filter is 1.065 times larger, and using it instead lowers $G$ by \SI{0.27}{dB}, which we take as a systematic uncertainty. A cable connecting the two switches then gives the input attenuation from the source to the device ports, $A_{\rm in}=41.6$\,dB.

Each measured trace is $M_{ij}=c_{ij}(\omega)S_{ij}$, where the path factor $c_{ij}=a_jg_i$ is the product of the complex input factor $a_j$ from the source to port $j$ and the output factor $g_i$ from port $i$ to the receiver, so that $c_{11}c_{22}=c_{12}c_{21}$. The linear fit determines the four path factors together with the device response and places the scattering parameters on an absolute scale at the chip ports. The fit imposes $c_{11}c_{22}=c_{12}c_{21}$ on the mean levels of the path factors. If the two ports are in addition assumed to have equal input attenuation, the transmission path factors are reproduced only to within 0.7\,dB, which we take as the systematic uncertainty of the absolute transmission levels. The contrast is a ratio of two transmissions measured through the same lines and is affected only through the ratio of the transmission path factors. Two independent measurements fix the relative drive of the two ports. The first is this linear calibration, whose transmission path factors require the input attenuation of port~1 to exceed that of port~2 by 0.5\,dB, consistent within the closure uncertainty with a separate measurement of the two port branches, which differ by 0.1--\SI{0.2}{dB}. The second is the power-rescaling collapse of the nonlinear maps, which needs no model and gives a larger effective offset of -1.4\,dB. The collapse measures the ratio of drive strengths $\gamma_1P_1/\gamma_2P_2$ rather than the delivered powers alone, so the difference of about \SI{1}{dB} may come from a slightly smaller coupling ratio at high stored energy or from a loss in the port-1 branch that the linear calibration does not capture. We use the collapse value in the nonlinear model. Pairing traces at equal chip-port powers for any offset between the two values changes the peak contrast by less than \SI{1}{dB} (Supplementary Note~10), which bounds the effect of this uncertainty on the main result. Mode~1 was calibrated in the same way with its own linear fit (Supplementary Note~12). Its absolute drive power is less certain, and the comparison of the two modes in Fig.~\ref{fig:parity}e uses the measured Kerr shift as the measure of the drive.

\subsection*{Coupled-mode model}
We describe the device with temporal coupled-mode theory\cite{Fan2003,Suh2004}. The resonator amplitude $a$, normalized such that $|a|^2$ is the stored energy, obeys $\dot a=[i\omega_c(|a|^2)-\gamma]a+\bm k^{\rm T}\bm S^+$ and radiates into the outgoing waves, $\bm S^-=C\bm S^++\bm k a$. Here $\omega_c(|a|^2)$ is the energy-dependent resonance frequency, $\bm S^+$ and $\bm S^-$ are the vectors of incoming and outgoing wave amplitudes, normalized so that $|S_j^\pm|^2$ is the power in channel $j$, and $\bm k$ is the vector of coupling amplitudes. The background matrix $C$, the scattering matrix of the line without the resonator, is symmetric and unitary, $|k_j|^2=2\gamma_j$, and energy conservation with time-reversal symmetry imposes $C\bm k^*=-\bm k$. Besides the two signal ports we include a third channel p, so that $\gamma=\gamma_1+\gamma_2+\gamma_{\rm p}+\gamma_{\rm i}$. We parametrize the background as $C=O\,{\rm diag}(e^{i\phi_1},e^{i\phi_2},e^{i\phi_{\rm p}})\,O^{\rm T}$ with a real rotation $O$, and every coupling vector that satisfies the constraint as $\bm k=O\,(ie^{i\phi_1/2}s_1,ie^{i\phi_2/2}s_2,ie^{i\phi_{\rm p}/2}s_{\rm p})^{\rm T}$ with real coupling strengths $s_m$ and background phases $\phi_m$ ($m=1,2,{\rm p}$). The transmission then takes the Fano form $S_{21}=t_{\rm b}[1+Z\gamma/(i(\omega-\omega_c)+\gamma)]$ with $Z=k_1k_2/t_{\rm b}\gamma$, which gives Eq.~\eqref{eq:fano} with $\varepsilon=1+{\rm Re}\,Z$ and $c={\rm Im}\,Z$. Its extrema $|t_{\rm b}|^2\lambda_\pm$ follow from setting the derivative of Eq.~\eqref{eq:fano} with respect to $y$ to zero (Supplementary Note~4). A two-port model has a $2\times2$ background and no channel p ($C_{1{\rm p}}=C_{2{\rm p}}=k_{\rm p}=0$). We fitted it with its own parametrization, and it gives ${\rm Re}\,Z=-(\gamma_1+\gamma_2)/\gamma$ and $\varepsilon=\gamma_{\rm i}/\gamma$, because energy conservation and time-reversal symmetry fix the relative phase of the direct and resonant paths, so that only the internal loss prevents a perfect zero. The closest approach of the transmission circle to the origin is $\varepsilon/\sqrt{1+c^2}$ in units of the direct path. The ceiling of Eq.~\eqref{eq:ceiling} is an upper bound, because reaching it requires both operating states to be accessible at the same frequency and power. Each path factor describes the gain and phase of the cables, circulators and amplifiers across the \SI{80}{MHz} window and is modelled as an electrical delay times a third-order complex polynomial in frequency, obtained by linear least squares for every set of physical parameters, with the mean levels constrained by $c_{11}c_{22}=c_{12}c_{21}$. These smooth functions vary slowly over the window and cannot reproduce the resonance, which is about \SI{10}{MHz} wide, so they do not absorb the device response. We fitted the average of the repeated traces recorded at \SI{-30}{dBm} (2000 frequencies for each of the four traces, 8000 complex points), weighting each point by its measured noise, from 60 random starting points. All starts that reached the global minimum gave the same parameters, which are $\omega_0/2\pi=10.0494$\,GHz, $|t_{\rm b}|^2=-14.6$\,dB, $Z=-0.941+3.25i$, and loaded, external and internal quality factors of 940 / 1370 / 3640, where the external quality factor refers to the combined coupling to the two signal ports. To test the third channel, we withheld one reflection trace at a time and predicted it from the other three. The three-channel model reproduces the shallow port-2 dip to within \SI{0.1}{dB} and the deep port-1 dip to within \SI{5}{dB}, whereas the two-port model misses them by \SI{8}{dB} and \SI{23}{dB} (Supplementary Note~5). The nearest line of perfect zeros (Fig.~\ref{fig:device}f) corresponds to a coupling ratio of 2.44 instead of 2.41, a leakage rate of \SI{0.27}{MHz} instead of 0.29\,MHz and a background transmission \SI{0.4}{dB} higher (Supplementary Note~7). The derivations are given in Supplementary Note~4.

\subsection*{Nonlinear model and fitting}
The Kerr shift $u=\chi|a|^2$ lowers the resonance, $\omega_c=\omega_0-u$. For a drive at port $j$ with detuning $\Delta=\omega-\omega_0$ the steady state obeys $u[(\Delta+u)^2+\gamma(u)^2]=\eta_jP_j$ with $\eta_j=2\gamma_j\chi$, where $\gamma(u)$ is the total decay rate including the energy-dependent internal loss defined below. The drive enters only through $\gamma_jP_j$, which gives the power-rescaling symmetry $S_{21}(\omega,P)=S_{12}(\omega,\rho P)$ and bistability for detunings beyond $\Delta_{\rm c}=-\sqrt3\gamma$\cite{Siddiqi2005}. To avoid overfitting we used a minimal model. All linear parameters are fixed at the values of the low-power fit. The only additions are the Kerr shift and a nonlinear internal loss, $\gamma_{\rm i}(u)=\gamma_{\rm i}+\beta u$, a known feature of strongly driven NbN resonators\cite{Abdo2006,Swenson2013}. The ratio of the drive strengths at the two ports is fixed by the measured power-rescaling shift (Supplementary Note~15). We express the Kerr coefficient through the Kerr shift per photon $K$, which is negative for a downward shift, as $\chi=-K/\hbar\omega_0$, where $\hbar$ is the reduced Planck constant, so that $\eta_jP_j=-2\gamma_jKP_j/\hbar\omega_0$. Only two parameters are fitted, $K$ and the dimensionless nonlinear loss coefficient $\beta$. The model is stationary, so $K$ is an effective coefficient that also contains the slow contribution identified in the pulsed measurements. The physical root of the steady-state equation was selected by adiabatic continuation along the measured sweep direction. Each frequency point is held for about \SI{1}{ms}, much longer than the microsecond response found in the pulsed measurements, so the maps sample the stationary response. We fitted the calibrated magnitudes of all four scattering parameters for both sweep directions with a soft-$\ell_1$ loss, which is quadratic for small residuals and linear for large ones and so limits the weight of outliers, excluding points within \SI{1.2}{MHz} of the transient frequencies and the hysteretic regions of the upward sweeps, where switching between the metastable branches is sensitive to noise and to source transients\cite{Siddiqi2005}. The fit gives $K/2\pi=-0.75$\,Hz per photon and $\beta=0.187$, with a median absolute deviation of 0.13\,dB. Leaving the attenuator step free instead lowers the cost by less than \SI{3}{\percent}, so the fit does not need it. The model is evaluated on a fine grid of \SI{0.05}{dB} in power and \SI{20}{kHz} in frequency for all model maps and curves. The low-power slopes of the reflection dips for drive at port~1 and port~2, -1.01 and -1.40\,Hz per photon, give $K/2\pi=-1.2\pm0.2$\,Hz per photon, where the uncertainty is half their difference. The \SI{40}{\percent} difference between the two determinations corresponds to about \SI{2}{dB} in delivered power. It affects only the absolute photon-number scale and the value of $K$, not the contrast, the switching edges or the power-rescaling collapse, which are all compared on the measured source-power axis. Depending on which value is used, the resonator stores between $2.6\times10^{7}$ and $3.6\times10^{7}$ photons at the contrast maximum. Without the nonlinear loss the model predicts the onset of bistability at far too low a power and too much dragging. The model does not reproduce the deepening of the transmission zero at high power, as discussed in the main text. Supplementary Note~6 compares it with a pure Kerr model and with a phenomenological extension.

\subsection*{Layout model of the parasitic channel}
A coplanar line supports, besides its signal mode, a slot-line mode in which the two ground planes oscillate against each other, and a bridge that shorts the centre conductor to one ground plane necessarily launches both. The layout model, implemented in custom Python code (Code availability), tests whether the bridge could also open the leakage channel. The feed line near the resonator was modelled as two uniform three-conductor cross-sections, the coplanar line and the resonator pocket, with zero-thickness NbN conductors on silicon between a conducting backplane and lid. We used $\varepsilon_{\rm r}=11.9$ and checked that the cryogenic value of 11.45 (ref.~\citenum{Krupka2006}) changes the results by less than \SI{0.15}{dB}. Capacitance matrices were obtained from a finite-volume solution of Laplace's equation, and inductance matrices from a method-of-moments solution of the magnetostatic problem that includes a sheet kinetic inductance of \SI{58}{pH/sq}, the value that matches the measured resonance frequency. This value is taken from the measured mode frequencies, so the model is not fully independent of the measurement, but none of its parameters is adjusted to the leakage. The sections and the bridge, a shunt inductance of \SI{0.38}{nH} between the centre conductor and one ground plane, were combined with multiconductor transmission-line theory\cite{Paul2008} into a scattering matrix for the signal, slot-line and common modes. The coupling of each resonator mode to each channel was obtained by reciprocity, as the overlap of its voltage profile along the meander with the voltage that the corresponding incoming wave produces at the strip. Including the measured rows of wire bonds leaves the conversion, the coupling asymmetry and the Fano orientations unchanged, but the bonds reflect 80--\SI{90}{\percent} of the slot-line amplitude. The model does not reproduce the leakage rate $\gamma_{\rm p}$ of the resonator or the split of the leakage between the two ports, which depend on details it does not include, such as the DC line, the bends of the feed line and the package (Supplementary Note~14). In the model, a second bridge across the other slot reduces the conversion from \SI{-10.3}{dB} to \SI{-22}{dB}, and a fully symmetric junction reduces it to \SI{-57}{dB}.

\section*{Data availability}
The data underlying all figures are available from the corresponding author upon reasonable request.

\section*{Code availability}
The code of the layout model is provided as Supplementary Software. The analysis and fitting code is available from the corresponding author upon reasonable request.

\section*{Acknowledgements}
This work was supported by a Natural Sciences and Engineering Research Council of Canada (NSERC) Discovery Grant and by the University of Calgary. A.A. was supported by the Department of Defense and the National Science Foundation.

\section*{Author contributions}
B.M. designed the chip, A.M. fabricated the device, and B.M. performed the measurements. S.B. and B.M. analysed the data. S.B. conceived the idea. A.A. contributed to the analysis of the results and their physical interpretation. All authors contributed to writing the paper.

\section*{Competing interests}
The authors declare no competing interests.

\end{document}